\documentclass[aps,prl,amsmath,amssymb,floatfix,reprint,superscriptaddress]{revtex4-2}

\usepackage{graphicx}
\usepackage{color}
\usepackage{microtype}
\usepackage{bm}
\usepackage{hyperref}
\hypersetup{colorlinks,allcolors=blue}

\def\equationautorefname#1#2\null{Eq.#1(#2\null)}

\usepackage{comment}
\usepackage{color,soul}

\usepackage[caption=false]{subfig}

\begin{document}

\title{Kubo-Anderson theory of polariton lineshape}

\author{Cl\`audia Climent}
\email{ccliment@sas.upenn.edu}
\affiliation{Department of Chemistry, University of Pennsylvania, Philadelphia, Pennsylvania 19104, USA}
\author{Joseph E. Subotnik}
\affiliation{Department of Chemistry, University of Pennsylvania, Philadelphia, Pennsylvania 19104, USA}
\author{Abraham Nitzan}
\affiliation{Department of Chemistry, University of Pennsylvania, Philadelphia, Pennsylvania 19104, USA}
\affiliation{School of Chemistry, Tel Aviv University, Tel Aviv 69978, Israel}

\begin{abstract}
We apply the Kubo-Anderson stochastic theory of molecular spectral lineshape to the case of polaritons formed in the collective strong coupling regime. We investigate both the fast and slow limits of the random frequency modulation of the emitter as well as the intermediate regime and show how the interplay between the characteristic timescales of the cavity and the molecular disorder is expressed in the observed polaritons lineshapes. The analytical solution obtained for the slow limit is valid for any ratio between the inhomogeneous broadening of the molecules and the Rabi splitting, especially relevant for molecular polaritons where these two quantities can be of the same order of magnitude.
\end{abstract}

\maketitle

%===================================================================================
\emph{Introduction.}---
When the interaction between a photon and an electronic/vibrational transition is strong enough such that their rate of energy exchange exceeds that of their respective losses, new hybrid light-matter states known as polaritons are formed \cite{laussy}.
One of the most interesting features of this strong light-matter coupling regime is the collective interaction of an ensemble of emitters with the electromagnetic field in optical cavities. 
Spectroscopically, this translates in an energetic (Rabi) splitting between the two polariton modes that scales with the square root of the number of emitters \cite{abe-annrev}. 
This collective response and the concept of a polariton as a coherent superposition of states with many different excited molecules naturally raises a question about the possible role of disorder. 
An interesting spectroscopic as well as numerical observation is that, in the presence of static disorder, and for a sufficiently large Rabi splitting, the polariton linewidth does not inherit the inhomogeneous broadening of the cavity-free emitters \cite{1995-yamamoto,1996-houdre,1998-houdre,simpkins-2015}. Instead, the polariton broadening is exclusively due to the homogeneous linewidth of both of its constituents, the cavity and emitter resonances. 

Several works have investigated this subject and closely related matters in the past, mostly within the context of semiconductor microcavities 
\cite{1996-houdre,1996-savona,1996-whittaker1,1996-whittaker2,1996-yamamoto,1997-savona1,1997-savona2,1998-houdre,1998-kavokin1,1998-kavokin2,1998-whittaker,2000-hvam,2001-agranovich,2003-agranovich,marchetti,2017-carusotto}.
Typically, numerical simulations were carried out to investigate the effect of static disorder on the polariton linewidth, while the effect of homogeneous broadening is usually treated phenomenologically.
Despite recent interest in the role of disorder in polaritonic phenomena 
\cite{michetti1,michetti2,michettichapter,johannes,scholes,genes,schachenmayer,chuntonov,cao1,cao2,ribeiro,herrera,guillaume,joel1,joel2,tgera1,tgera2,zeb,theta,zeyu,musser1,musser2,musser3,gerrit,climent}, 
especially within the context of molecular polaritons (where molecular transition bands are quite broad in comparison to atomic systems or semiconductors), an analytic theory capable of describing the effect of both static and dynamic disorder in the polariton lineshape is missing. 
In this Letter, we address this point by extending the Kubo-Anderson theory of a stochastic molecular lineshape \cite{anderson1953,anderson1954,kubo1954,kubo1957,kubo1963,kubo1969} to the case of many molecules that respond collectively to an optical excitation and, via the same collective response, form polaritons when interacting with resonance cavity modes. 
Our theory yields simple analytical results in the slow and fast limits of the disorder dynamics, and can be evaluated numerically for the intermediate case. The lineshape expression we obtain for the slow limit is valid for any ratio between the inhomogeneous broadening of the molecules and the Rabi splitting, especially relevant for molecular polaritons where the broadening due to static disorder can be a significant fraction of the Rabi splitting. 

%===================================================================================
\emph{Kubo-Anderson theory of stochastic molecular lineshape.}---
The starting point of the Kubo-Anderson theory of stochastic lineshape is to model a molecular transition as a classical harmonic oscillator whose frequency randomly fluctuates about a central frequency $\omega_0$ due to the interaction with a thermal environment \cite{kubo1969}. The dynamics of such an oscillator is described by the following equation of motion.
\begin{equation}\label{eq:kuboeom}
    \dot{a} = -i(\omega_0 + \delta \omega(t)) a
\end{equation}
The main assumption of the model is that the stochastic time-dependent frequency fluctuation $\delta \omega (t)$ caused by environmental motions is a random stationary Gaussian process characterized by an average 
$\langle \delta \omega (t) \rangle = 0 $ 
and an autocorrelation function for which a common model is 
$\langle \delta \omega (t) \delta \omega (t + \tau) \rangle = \Omega^2 e^{-\tau/\tau_c} $ 
with a correlation time
$\tau_c = \frac{1}{\Omega^2}
\int_0^\infty d\tau \langle \delta \omega(t) \delta \omega(t + \tau) \rangle$
where  
$\Omega = \sqrt{\langle \delta \omega^2\rangle} $ 
is the amplitude of the random frequency modulations. 
The lineshape may then be obtained by calculating the Fourier transform of the autocorrelation function of the amplitude $a$, 
$I(\omega) = \int_{-\infty}^\infty dt \, e^{-i\omega t} \langle a^*(0) a(t) \rangle$, 
and different physical behaviors are encountered depending on the relative magnitude of the correlation time $\tau_c$ and the amplitude of the frequency modulations $\Omega$. Analytical solutions can be obtained in two extreme limits characterized by the magnitude of the dimensionless parameter $\alpha \equiv \tau_c \Omega$:

(i) $ \alpha \ll 1 $ represents the fast limit, that is, the situation where the dynamics of the environment is fast relative to that of the oscillator. A Lorentzian lineshape $I(\omega) = \frac{2 \Gamma}{\omega^2 + \Gamma^2}$ 
is obtained in this limit with a half-width half-maximum $\Gamma=\tau_c\Omega^2 = \alpha\Omega$ that can be much narrower than the amplitude of the frequency modulation, a phenomenon known as motional narrowing that has been extensively investigated in NMR spectroscopy \cite{motional-narrowing-NMR,book-nmr}. 
\footnote{Note that the term motional narrowing has lead to misleading interpretations in the past when describing narrowing of polariton linewidth in semiconductor microcavities \cite{1996-whittaker2, 1997-savona2, 1998-kavokin1}. Static disorder (and not fast disorder) appears to be responsible for such narrowing \cite{1998-houdre, 1996-houdre}.} 
Note that the fast limit of the Kubo-Anderson theory is equivalent to the Markovian Bloch-Redfield theory, where it becomes clear that the intrinsic relaxation of the system (with contributions from both population relaxation and pure dephasing) is responsible for the so-called homogeneous broadening. 

(ii) $ \alpha \gg 1 $ corresponds to the slow limit where the dynamics of the bath is slow compared to the inverse of the amplitude of the random frequency modulations. A Gaussian lineshape 
$I(\omega) = \sqrt{\frac{2\pi}{\Omega^2}}e^{-\frac{\omega^2}{2\Omega^2}}$ is obtained in this case, characterized by a width $\Omega$ whose inhomogeneous character stems from the fact that each oscillator in an ensemble will experience different frequency shifts because of the slow dynamics of the environment, i.e., every oscillator will experience a ``different" environment.

%===================================================================================
\emph{Model for polariton lineshape}---
The Kubo-Anderson solution of the lineshape of randomly modulated molecules treats a single molecule interacting with the radiation field and takes an average over an ensemble of such molecules. 
A naive extension to the molecule-in-cavity problem would be to consider an ensemble of systems, each comprising a single molecule and a cavity mode. 
However, such an extension of the original model would not be physical because a single-molecule is not sufficient to reach the strong coupling regime and thus such a model would not correspond to any realizable experimental situation.
By contrast, a physically sound extension of the Kubo-Anderson model consists of a cavity photon ($a_c$) coupled to $N$ molecules ($a_j$), essentially a Tavis-Cummings model \cite{Tavis1968} with modulated molecular transition frequencies. Note that the (static) disordered Tavis-Cummings model has recently been investigated via the Green's function approach \cite{tgera1, tgera2, zeb, cao2}. 
Here we follow Anderson and Kubo and investigate both the static and dynamic disorder cases and the transition between them. We represent the molecules by classical harmonic oscillators which, under driving by an incident radiation field $Fe^{-i\omega t}$ 
\footnote{We take $F=1$ in the calculations.}
evolve according to \footnote{In addressing the driven Tavis-Cummings model one may choose to couple the molecules or the cavity mode to the external radiation field and we take the latter as a more physically appealing choice. If one were to solve the original out-of-cavity Kubo-Anderson problem with this approach, the driving would obviously be on the molecule, 
$\dot{a}_j(t) = -i(\omega_0 + \delta \omega(t))a_j -\gamma a_j +iFe^{-i\omega t}$.}
\begin{equation}
\begin{split}
    \dot{a}_c(t) &= -i\omega_c a_c -iu\sum_j^N a_j - \kappa a_c + iFe^{-i\omega t } \\
    \dot{a}_j(t) &= -i(\omega_j + \delta \omega_j(t)) a_j -iu a_c - \gamma a_j \\
\end{split}
\end{equation}
where $\omega_c$ is the photon frequency, $\omega_j$ is the time-independent molecular transition frequency,  $\delta \omega_j(t)$ is the random frequency modulation of the molecular transition, $u$ is the single-molecule coupling strength, and $\kappa$ and $\gamma$ are the dampings of the photon and molecules, respectively. 
\footnote{The relaxation parameter $\gamma$ is needed to facilitate the steady state treatment that follows. It may be set to zero at the end of the calculation, or kept to reflect the finite lifetime of individual molecules. Collective (superradiant) relaxation, see e.g. \cite{theta,zeyu}, is disregarded in the present treatment.}
In the following we focus on the on-resonance situation with $\omega_0 \equiv \omega_c = \omega_j$, where the cavity response in the strong-coupling regime is characterized by two polariton peaks separated by the (collective) Rabi splitting $\Omega_R = 2\sqrt{N}u$.
In steady state, the solutions oscillate with the driving frequency, i.e., $a_c(t)=\bar{a}_c(t) e^{-i\omega t}, a_j(t)=\bar{a}_j(t) e^{-i\omega t}$, so the equations of motion become 
\begin{subequations}\label{eq:eqdot}
\begin{align}
    \dot{\bar{a}}_c(t) &= -i\bar{\omega}_0 \bar{a}_c -iu\sum_j^N \bar{a}_j - \kappa \bar{a}_c + iF \label{eq:dotac} \\ 
    \dot{\bar{a}}_j(t) &= -i(\bar{\omega}_0 + \delta \omega_j(t)) \bar{a}_j -iu \bar{a}_c - \gamma \bar{a}_j \label{eq:dotaj}
\end{align}
\end{subequations}
where we have defined $\bar{\omega}_0 \equiv \omega_0 - \omega$.
The average total energy of the system 
$\langle E(t) \rangle = \omega_0 \langle a^*_c a_c \rangle + \sum_j^N \omega_0 \langle a^*_j a_j \rangle $ 
\footnote{We have not included the $\delta \omega_j(t)$ contribution to the energy because these fluctuations are much smaller than $\omega_0$}
satisfies in steady-state
\begin{equation}
    \Big \langle \frac{dE}{dt} \Big \rangle
    = \Big \langle \frac{dE}{dt} \Big \rangle_{in} + \Big \langle \frac{dE}{dt} \Big \rangle_{out} = 0
\end{equation}
allowing to identify the pumping and damping contributions as
\begin{subequations}
\begin{align}
    \Big \langle \frac{dE}{dt} \Big \rangle_{in} &= 
    i(F \langle \bar{a}_c^* \rangle - F^* \langle \bar{a}_c \rangle) \label{eq:dEdt} \\
    \Big \langle \frac{dE}{dt} \Big \rangle_{out} &= 
    -2\kappa \langle |\bar{a}_c|^2 \rangle -2\gamma \sum_j^N \langle |\bar{a}_j|^2 \rangle
\end{align}
\end{subequations}
The absorption lineshape may be obtained by evaluating, for instance, \autoref{eq:dEdt}, as a function of the incident frequeny $\omega$. To this end we only need to find $ \langle \bar{a}_c \rangle$. For a single molecule outside the cavity, this approach leads to the familiar Kubo-Anderson result. \cite{suppmaterial}

We proceed by integrating \autoref{eq:dotaj}
\begin{equation}
\begin{split}
    \bar{a}_j(t) = 
    \bar{a}_j(t_0) e^{-i\bar{\omega}_0 (t-t_0) 
    - \gamma(t-t_0) -i\int_{t_0}^t \delta \omega_j (t'') \,dt''} \\
    -iu \int_{t_0}^t dt' \, 
    e^{ -i\bar{\omega}_0 (t-t') - \gamma(t-t') 
    -i \int_{t'}^{t} \delta \omega_j (t'') \, dt''}
    \bar{a}_c(t')
\end{split}
\end{equation}
where the first term corresponds to the transient and only the second contributes to the steady-state solution. Using it in \autoref{eq:dotac} we find
\begin{equation}
\begin{split}
    &\dot{\bar{a}}_c(t) = -(i\bar{\omega}_0 + \kappa)\bar{a}_c + iF \\
    &-u^2 \sum_j^N \int_{-\infty}^t dt' \, e^{-i\bar{\omega}_0(t-t') -\gamma (t-t')}
    e^{-i\int_{t'}^t \delta \omega_j(t'')dt''} \bar{a}_c(t')
\end{split}
\end{equation}
Also, at steady state (i.e., $t \rightarrow \infty$), $\bar{a}_c(t')$ can be taken outside the integral by the following argument: When $t'$ is large (i.e., $t' \rightarrow t$), $\bar{a}_c(t')$ is a constant, while when $t'$ is small (i.e., $t' \rightarrow -\infty$), the term vanishes. This leads to 
\begin{equation}\label{eq:dotacsol}
\begin{split}
    &\dot{\bar{a}}_c(t) = -(i\bar{\omega}_0 + \kappa)\bar{a}_c + iF \\
    &-u^2 \bar{a}_c \int_{-\infty}^t dt' \, e^{-i\bar{\omega}_0(t-t') -\gamma (t-t')}
    \sum_j^N e^{-i\int_{t'}^t \delta \omega_j(t'')dt''}
\end{split}
\end{equation}
Irrespective of the timescale of the frequency modulation, 
\begin{equation}\label{eq:average}
    \sum_j^N e^{-i\int_{t'}^t \delta \omega_j(t'')dt''} \approx N \big \langle e^{-i\int_{t'}^t \delta \omega_j(t'')dt''} \big \rangle
\end{equation}
is a reasonable approximation for large $N$, leading to 
\begin{equation}\label{eq:dotacsolsum}
\begin{split}
    &\dot{\bar{a}}_c(t) = -(i\bar{\omega}_0 + \kappa)\bar{a}_c + iF \\
    &-Nu^2 \bar{a}_c \int_{-\infty}^t dt' \, e^{-i\bar{\omega}_0(t-t') -\gamma (t-t')}
    \Big \langle e^{-i\int_{t'}^t \delta \omega_j(t'')dt''} \Big \rangle ,
\end{split}
\end{equation}
and because at steady state $\langle \dot{\bar{a}}_c \rangle = 0$, it follows that
\begin{equation}\label{eq:acgen}
\begin{split}
    &\langle \bar{a}_c \rangle = \\
    &\frac{iF}{i\bar{\omega}_0 + \kappa + 
    Nu^2 \int_{-\infty}^t dt' \, e^{-i\bar{\omega}_0(t-t')-\gamma (t-t')}\Big \langle e^{-i\int_{t'}^t \delta \omega_j(t'')dt''} \Big \rangle}
\end{split}
\end{equation}
Knowing that $\Big \langle e^{-i\int_{t'}^t \delta \omega_j(t'')dt''} \Big \rangle$ is a function of $t-t'$ in the present model \cite{suppmaterial}, we have 
\begin{equation}\label{eq:ac}
    \langle \bar{a}_c \rangle = \frac{iF}{i\bar{\omega}_0 + \kappa + 
    Nu^2 \int_0^\infty dt \, e^{-i\bar{\omega}_0 t -\gamma t} \phi(t)}
\end{equation}
with
\begin{equation}
\phi(t) = \Big \langle e^{i\int_0^t \delta \omega_j(t')dt'} \Big \rangle
\end{equation}
which can be evaluated as in the Kubo-Anderson work \cite{anderson1954,kubo1954}.
Eq. (\ref{eq:ac}) will be our starting point to investigate the two limiting cases where the molecular transition is either homogeneously or inhomogeneously broadened, as well as the intermediate regime.

As a final remark, note that in order to obtain a general handleable expression like \autoref{eq:ac} leading to analytical results for both the fast and slow limits (vide infra), it was crucial to use \autoref{eq:average} before taking ensemble averages. If otherwise we had set $\dot{\bar{a}}_c=0$ in \autoref{eq:dotacsol} for the slow case and taken the average over realizations before using \autoref{eq:average}, we would have obtained a far more complex expression for $\langle \bar{a}_c \rangle$ in the slow limit that would have required some approximation in order to be solved.

%===================================================================================
\emph{Fast limit.}---
In the fast modulation limit $\phi(t)=e^{-\Gamma t}$ \cite{anderson1954,kubo1954}. \autoref{eq:ac} then leads to 
\begin{equation}
    \langle \bar{a}_c \rangle = \dfrac{iF}{i\bar{\omega}_0 + \kappa + 
    \dfrac{Nu^2}{i\bar{\omega}_0 + \gamma_m}}
\end{equation}
where $\gamma_m = \gamma + \Gamma$ is the total relaxation rate of the molecule, with pure dephasing rate $\Gamma$ and lifetime broadening $\gamma$. 
From the driving term in \autoref{eq:dEdt} we find the spectrum to have a Lorentzian profile 
\begin{equation}\label{eq:Iwfast}
\begin{split}
    I(\omega) &= |F|^2 
    \frac{2\kappa|i\bar{\omega}_0 + \gamma_m|^2 + 2\gamma_m Nu^2}
    {|(i\bar{\omega}_0 + \kappa)(i\bar{\omega}_0 + \gamma_m) + Nu^2|^2}
\end{split}
\end{equation}
with the poles located at 
\begin{equation}\label{eq:poles}
    \omega = \omega_0 -\frac{i}{2}(\gamma_m + \kappa) \pm 
    \sqrt{Nu^2 - \Big(\frac{\gamma_m-\kappa}{2}\Big)^2}
\end{equation}
By assuming 
$\sqrt{N}u \gg (\gamma_m - \kappa)/2 $, which is reasonable since we are interested in the collective strong coupling regime, we find the two polariton peaks at
$\omega_0 \pm \sqrt{N}u -\frac{i}{2}(\gamma_m + \kappa)$, where they are split by the collective Rabi splitting $\Omega_R$ and each peak inherits half of the original broadening of the cavity and molecular resonances. 
In particular when $\gamma_m=\kappa$, \autoref{eq:Iwfast} becomes 
\begin{equation}
    I(\omega) = |F|^2 \Bigg( \frac{\gamma_m}{(\bar{\omega}_0 - \sqrt{N}u)^2 + \gamma_m^2}
    + \frac{\gamma_m}{(\bar{\omega}_0 + \sqrt{N}u)^2 + \gamma_m^2} \Bigg)
\end{equation}

%===================================================================================
\emph{Slow limit.}---
A shortcut to explore the effect of static disorder on polariton broadening is to use the fact that in the slow modulation limit, $\delta \omega_j$ are time-independent. Hence, $\dot{\bar{a}}_c=0$ and $\dot{\bar{a}}_j=0$, so from \autoref{eq:eqdot} we have 
\begin{equation}
    \bar{a}_c =
    \dfrac{iF}{i\bar{\omega}_0 + \kappa + \sum_j^N 
    \dfrac{u^2}{i(\bar{\omega}_0 + \delta\omega_j) + \gamma}} 
\end{equation}
and the lineshape for a given realization (using \autoref{eq:dEdt} without the ensemble average) is 
\begin{equation}
    I(\omega) = |F|^2 \frac{2\kappa}
    {\Big(\bar{\omega}_0 - \sum_j^N \dfrac{u^2}{\bar{\omega}_0 + \delta \omega_j}\Big)^2 + \kappa^2}
\end{equation}
where for the sake of simplicity we have neglected homogeneous broadening ($\gamma=0$). 
To make progress we expand the denominator for $\delta \omega_j/\bar{\omega}_0 \ll 1$ (which is satisfied in the vicinity of the polariton frequencies where $\bar{\omega}_0 \sim \pm \sqrt{N}u$ for strong enough coupling) and find
\begin{equation}\label{eq:Iwslowapprox}
    I(\omega) = |F|^2 \frac{2\kappa}
    {\Big(\bar{\omega}_0 - \dfrac{Nu^2}{\bar{\omega}_0}
    +\dfrac{u^2}{\bar{\omega}_0^2}W_N \Big)^2 + \kappa^2}
\end{equation}
where we have defined $W_N \equiv \sum_j^N \delta \omega_j$. This random number is characterized by the average $\langle W_N \rangle = 0 $ and variance 
$\langle \delta W_N^2 \rangle = N \langle \delta \omega_j^2 \rangle $.
To understand the effect of static disorder on the position and broadening of the peaks we must analyze the zeros of the following term.
\begin{equation}
   \bar{\omega}_0 - \dfrac{Nu^2}{\bar{\omega}_0} + \dfrac{u^2}{\bar{\omega}_0^2}W_N = 0
\end{equation}
We proceed to solve the above expression for 
$ \bar{\omega}_0 = \bar{\omega}_0^0 + \Delta \bar{\omega}_0 $ where 
$\bar{\omega}_0^0 = \pm \sqrt{N}u$ and the effect of static disorder is contained in 
$\Delta \bar{\omega}_0$. To lowest order in $\Delta \bar{\omega}_0$ we find 
\begin{equation}
    \Delta \bar{\omega}_0 = - \frac{u^2}{{2\bar{\omega}_0^0}^2} W_N
\end{equation}
The variance of this term represents the effect that static disorder has on the broadening and is given by
\begin{equation}\label{eq:scaling}
    \langle \delta \Delta \bar{\omega}_0 ^2 \rangle = 
    \Big( \frac{u^2}{{2\bar{\omega}_0^0}^2} \Big)^2 \langle \delta W_N^2 \rangle
    \sim \frac{\langle \delta \omega_j^2 \rangle}{N}
\end{equation}
We see that $\langle \delta \Delta \bar{\omega}_0 ^2 \rangle^{1/2}$ scales like $1/\sqrt{N}$, confirming that in the collective regime, polaritons are immune to broadening due to static disorder for sufficiently large Rabi splitting. Note that the $1/\sqrt{N}$ scaling result was recently obtained with a more involved treatment \cite{tgera1}. 

A general expression for the lineshape can be obtained using $\phi(t) = e^{-\frac{1}{2}\Omega^2t^2}$  \cite{anderson1954,kubo1954} in \autoref{eq:ac} \cite{suppmaterial} which leads to 
\begin{equation}\label{eq:Iwslowgen}
    I(\omega) = |F|^2 \frac{2(\kappa + \tilde{\gamma})}
    {(\bar{\omega}_0 + \Delta)^2 +
    (\kappa + \tilde{\gamma})^2}
\end{equation}
where $ \tilde{\gamma}(\omega) \equiv Nu^2\sqrt{\frac{\pi}{2\Omega^2}} e^{-\frac{\bar{\omega}_0^2}{2\Omega^2}} $ 
and 
$ \Delta(\omega) \equiv - \tilde{\gamma} \text{Erfi}\big[ \frac{\bar{\omega}_0}{\sqrt{2\Omega^2}} \big] $, with $\text{Erfi}$ denoting the imaginary error function, where for simplicity we have disregarded the intrinsic homogeneous broadening $\gamma$. This lineshape is Lorentzian and in addition to the cavity broadening $\kappa$ there is one of molecular origin $\tilde{\gamma}$. Note that \autoref{eq:Iwslowgen} is valid for any ratio of the Rabi splitting and inhomogeneous broadening, $\Omega_R/\Omega$, and therefore can describe the lineshape of molecular polaritons for the important and common case that the inhomogeneous broadening of the molecular species is a considerable fraction of the Rabi splitting. In the limit when $\Omega_R/\Omega \gg 1$, $\tilde{\gamma}$ vanishes at the polariton frequencies and only the cavity broadening remains \cite{suppmaterial}, in agreement with \autoref{eq:Iwslowapprox} and \autoref{eq:scaling}.

\begin{figure}[h]
\includegraphics[width=\linewidth]{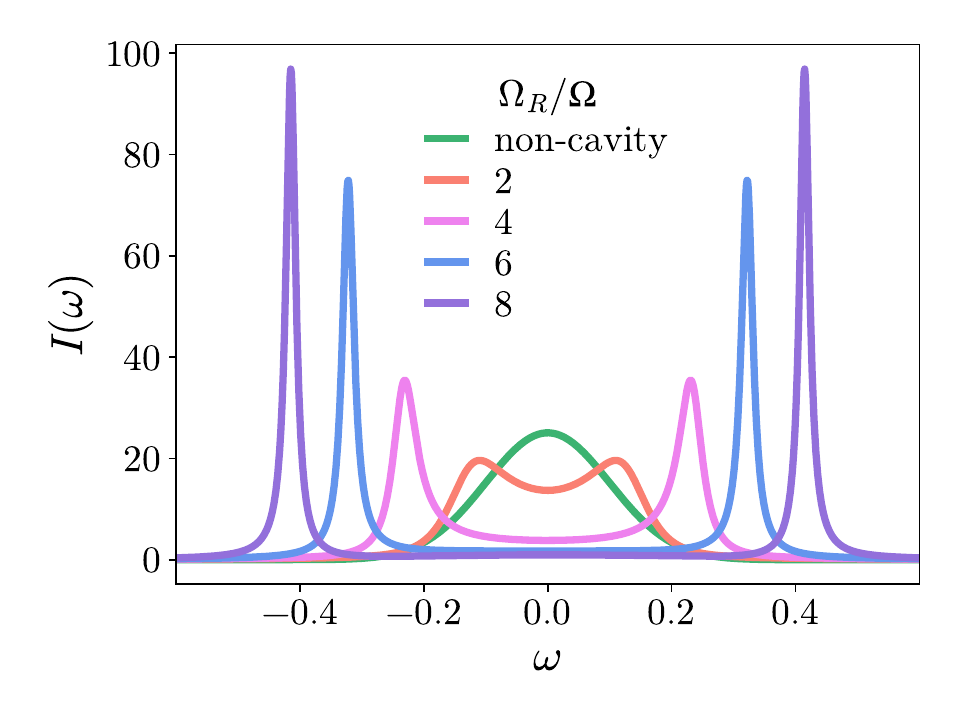}
\caption{(a) Analytical spectrum for the slow modulation limit in the non-cavity case \cite{suppmaterial} and for the cavity case (\autoref{eq:Iwslowgen}) with increasing ratio between the Rabi splitting and the amplitude of the random frequency modulation,  $\Omega_R/\Omega$, where $\Omega_R= 2\sqrt{N}u $.
Parameters: $\omega_0 = 0$, $\kappa=0.02$, $\gamma=0$, $\Omega=0.1$, $\alpha = 50$.}\label{fig-slow}
\end{figure}

In \autoref{fig-slow} we plot analytical results for the spectrum in the slow modulation limit with varying degree of Rabi splitting with respect to the amplitude of the random frequency modulation, $\Omega_R/\Omega$. We see that while outside the cavity the spectrum has a Gaussian profile, the polariton lineshape in the strong coupling regime is much narrower, with a Lorentzian lineshape whose width is reduced as the Rabi splitting increases relative to the homogeneous broadening. Note that for $\Omega_R/\Omega \approx 2$ the two polariton peaks are already visible despite the Rabi splitting being only twice the inhomogeneous broadening. This is a common situation in molecular polaritons, for instance, in \cite{karl} the molecular band had a Gaussian-like profile with a FWHM$\sim 530$ meV, which corresponds to $\Omega \sim 300$ meV, and the Rabi splitting was $\sim 600$ meV. Also, note that for $\Omega_R/\Omega \approx 4$ (pink line) the narrowing in the lineshape is already quite noticeable. We mention in passing that for small $\Omega_R/\Omega$, the two polariton bands arise from many eigenstates with a small cavity photon contribution, and not two clean polaritons, as already extensively discussed in the literature \cite{michetti1,michettichapter,gerrit,karl}. 

\begin{figure}[h]
\includegraphics[width=0.8\columnwidth]{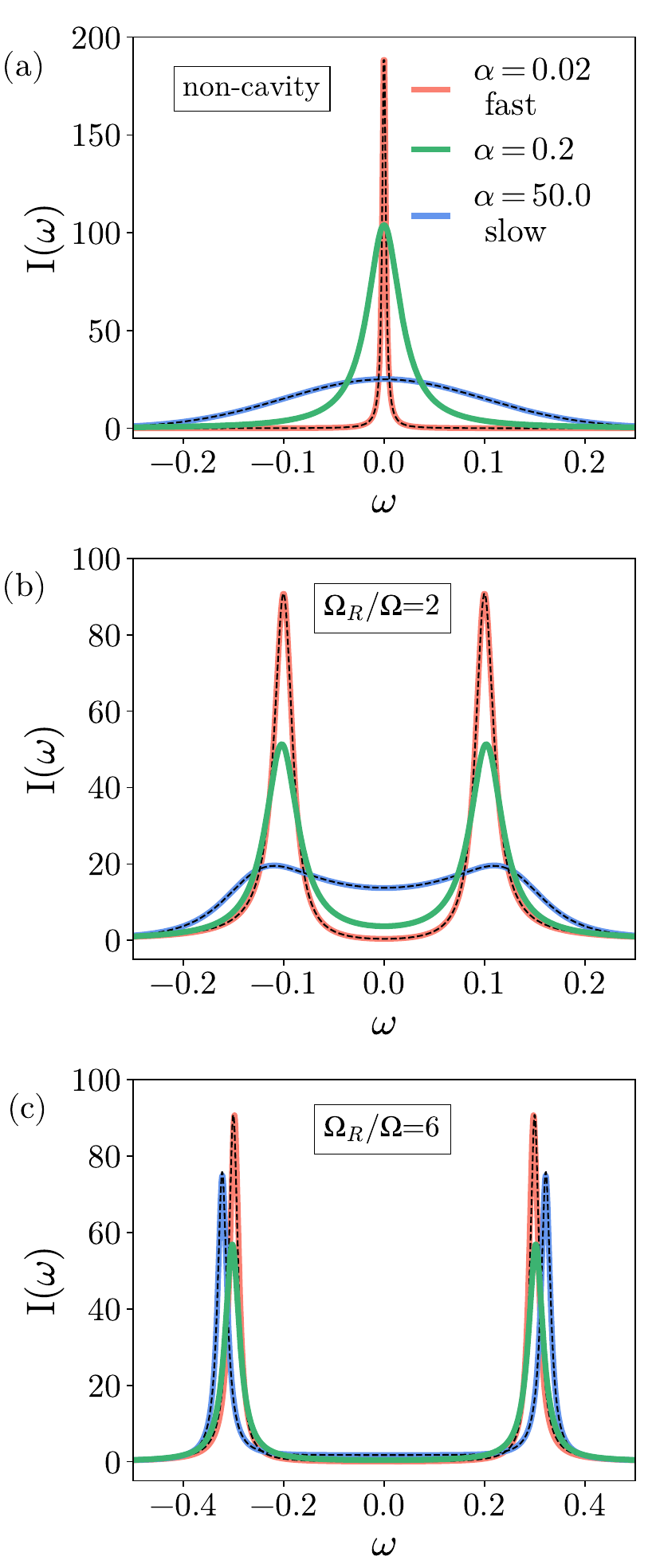}
\caption{Calculated spectrum for (a) non-cavity, and cavity cases (b) $\Omega_R/\Omega=2$, (c) $\Omega_R/\Omega=6$ for the fast ($\alpha=0.02$, red), intermediate ($\alpha=0.2$, green) and slow ($\alpha=50.0$, blue) modulations limits. 
The analytical spectrum for the fast and slow limits is also shown in black dashed lines, overlapping with the numerical results. 
Note that the non-cavity Lorentzian for the fast limit in (a) has been multiplied by a factor of 0.05. Parameters: $\omega_0 = 0$, $\kappa=0.02$, $\gamma=0$, $\Omega=0.1$. }\label{fig-slow-fast}
\end{figure}

%===================================================================================
\emph{Intermediate regime.}---
We here explore the intermediate regime where the correlation time of the random frequency modulations is comparable to the inverse of their amplitude, i.e., $\alpha \equiv \tau_c\Omega \approx 1$. To investigate the transition between the slow and fast limits we numerically calculate the spectrum with \autoref{eq:dEdt} and \autoref{eq:ac}. In this general case the ensemble averaged quantity is given by \cite{kubo1954,abebook}
\begin{equation}\label{eq:general}
    \phi(t) = 
    \exp{\Big[-\alpha^2 \Big( \frac{t}{\tau_c} - 1 + e^{-t/\tau_c} \Big) \Big]}\\
\end{equation}
with $\alpha \equiv \tau_c \Omega$ determining the transition between the fast ($\alpha \ll 1 $) and slow ($\alpha \gg 1 $) limits. 
In \autoref{fig-slow-fast} we plot numerical results for varying parameter $\alpha$  that controls the timescale of the frequency modulations relative to their amplitude. In all cases the spectrum smoothly transitions from the dynamic to the static disorder limit. For the two limits, $\alpha=0.02$ and $\alpha=50.0$, we also plot (in black dashed lines) the analytical spectrum which overlaps with the numerical results. While outside the cavity the lineshape is very different in the fast and slow limits (Lorentzian vs Gaussian), such a difference is reduced inside the cavity as $\Omega_R/\Omega$ increases. For instance, in \autoref{fig-slow-fast}(c), the lineshape is pretty narrow regardless of the $\alpha$ parameter.
Also, in these plots we can see that the frequency of the polariton peaks in the slow limit is slightly larger than that in the fast limit ($\pm 2\sqrt{N}u$), reflecting the effect of $\Delta$ in \autoref{eq:Iwslowgen}. Only when $\Omega_R \gg \Omega$ do the polariton frequencies for the slow limit coincide with those of the fast limit. 

%===================================================================================
\emph{Conclusions.}---
In this Letter we have applied Kubo-Anderson's theory of the stochastic lineshape to a model problem of coupled, driven and damped (classical) harmonic oscillators describing polaritons formed in the strong coupling regime. We have derived analytic expressions for the polariton lineshape in the limits of fast and slow disorder of the molecular transition frequency and numerically explored the intermediate regime as well. Our theory predicts that polaritons inherit half the original homogeneous broadening of the cavity and molecular resonance while static disorder does not contribute to their broadening for large enough Rabi splitting, in agreement with experimental observations and previous numerical calculations. Our results also provide an analytical expression for the polariton lineshape valid for any degree of static disorder relative to the Rabi Splitting, which is especially relevant within the context of molecular polaritons where the homogeneous broadening of the molecular transition can be a significant fraction of the Rabi splitting.
%===================================================================================
\begin{acknowledgments}
C.C. thanks the Vagelos Institute for Energy, Science, and Technology (VIEST) for a postdoctoral fellowship that initially supported this work. This project has received funding from the European Union’s Horizon 2020 research and innovation programme under the Marie Sklodowska-Curie grant agreement No 101029374. 
This work has been supported by the U.S. Department of
Energy, Office of Science, Office of Basic Energy Sciences,
under Award No. DE-SC0019397 (J.E.S.).
The research of A.N. was supported by the U.S. National
303 Science Foundation (grant no. CHE1953701).
\end{acknowledgments}
%===================================================================================
\bibliography{main}

%===================================================================================

\renewcommand{\theequation}{S\arabic{equation}}
\renewcommand{\thefigure}{S\arabic{figure}}

%===================================================================================

\cleardoublepage

\onecolumngrid
\begin{center}
	\textbf{SUPPLEMENTARY MATERIAL}
\end{center}

\twocolumngrid

%===================================================================================

\section{Retrieving the original Kubo-Anderson result}
In the original Kubo-Anderson model, the lineshape of the randomly modulated ostillator in \autoref{eq:kuboeom} was obtained by calculating the Fourier transform of the autocorrelation function of the amplitude $a$ \cite{kubo1957}. 
The physical justification behind this approach is that the Wiener–Khintchine theorem states that the spectral decomposition of the autocorrelation function of a random stationary process corresponds to its power spectrum. Also, the oscillator's position $x \sim (a + a^*)$, and so in thermal equilibrium, 
$\langle x(0) x(t) \rangle \sim \langle a^*(0) a(t) \rangle$ \cite{abebook}. 

In the main text we have used a different approach to extend the Kubo-Anderson theory to the case of polariton lineshape. In particular, we have obtained the spectral response by calculating the steady state of a collection of coupled, driven, and damped harmonic oscillators. 
In the following we show that this approach reduces to the original Kubo result for the case of a single molecule. 
The equation of motion for a driven and damped oscillator with a randomly modulated frequency is 
\begin{equation}
    \dot{a}(t) = -i(\omega_0 + \delta \omega(t))a -\gamma a +iFe^{-i\omega t}
\end{equation}
where we drop the $j$ index for the molecule in comparison to the main text. We are looking for the steady state, where the solutions will oscillate with the driving frequency. By setting 
$\bar{a}=a e^{-i\omega t}$ and $\bar{\omega}_0 \equiv \omega_0 - \omega$, the equation of motion reads
\begin{equation}\label{eqA:edo}
    \dot{\bar{a}}(t) = -i(\bar{\omega}_0 + \delta \omega(t))\bar{a} -\gamma\bar{a} + iF
\end{equation}
In the steady state, the average energy is constant with time
\begin{equation}\label{eqA:dEdt}
\Big\langle \frac{dE}{dt} \Big\rangle =
-2\gamma \langle |\bar{a}|^2 \rangle
+i(F \langle \bar{a}^* \rangle - F^* \langle \bar{a} \rangle)
=0
\end{equation}
and so the absorption spectrum may be calculated by evaluating the first term, which corresponds to the damping contribution, or the last two terms which correspond to the driving energy input.

The solution to the differential equation in \autoref{eqA:edo} is given by
\begin{equation}
\begin{split}
    \bar{a}(t) &= 
    \bar{a}(t_0) e^{-i\omega_0 (t-t_0) - \gamma(t-t_0) -i\int_{t_0}^t \delta \omega (t') \,dt'} \\
    &+ iF \int_{t_0}^t dt' \, 
    e^{ -i\bar{\omega}_0 (t-t') - \gamma(t-t') 
    -i \int_{t'}^{t} \delta \omega (t'') \, dt''}
\end{split}
\end{equation}
We are interested in the long-time dynamics, the stationary state. By taking $t_0=-\infty$, the first term does not contribute and the solution is given by the second term. Hence,  
\begin{equation}\label{eqA:bara}
    \langle \bar{a} \rangle = iF \int_{t_0}^t dt' \, 
    e^{-i\bar{\omega}_0 (t-t') - \gamma(t-t')} 
    \phi(t,t')
\end{equation}
where 
\begin{equation}
    \phi(t,t') \equiv
    \Big\langle e^{-i \int_{t'}^{t} \delta \omega (t'') \, dt''} \Big\rangle
\end{equation}
This average quantity may be calculated from cumulant averages as in \cite{kubo1962}, 
\begin{equation}
    \phi(t,t')
    = e^{-\frac{1}{2} \int_{t'}^t dt_1 \int_{t'}^t dt_2 \langle \delta\omega(t_1) \delta\omega(t_2) \rangle}
\end{equation}
In the present Kubo-Anderson model we have that 
$\langle \delta \omega (t) \delta \omega (t + \tau) \rangle = \Omega^2 s(\tau)$ where $\Omega = \sqrt{\langle \delta \omega^2\rangle} $ and $s(\tau)=s(-\tau)$. After some manipulation of the integral (like in \cite{abebook}) one can show that 
\begin{equation}
    \phi(t,t')  
    = e^{-\Omega^2 \int_0^{\tilde{t}} d\tau s(\tau) (\tilde{t}-\tau)}
\end{equation}
where $\tilde{t}=t-t'$. Taking $s(\tau)=e^{-\tau/\tau_c}$ \cite{kubo1963}, we obtain 
\begin{equation}\label{eqA:phi}
    \phi(\tilde{t}) =
    \exp{\Big[-\alpha^2 \Big( \frac{\tilde{t}}{\tau_c} - 1 + e^{-\tilde{t}/\tau_c} \Big) \Big]}\\
\end{equation}
which is a function of $t-t'$. After a change of variable in \autoref{eqA:bara} we get  
\begin{equation}\label{eqA:barafinal}
    \langle \bar{a} \rangle 
    = iF \int_0^\infty dt \, 
    e^{ -i\bar{\omega}_0 t- \gamma t} \phi(t)
\end{equation}
where
\begin{equation}
    \phi(t)=\Big \langle e^{i \int_0^t \delta \omega (t') \, dt'} \Big \rangle
\end{equation}
The fast ($\alpha \ll 1$) and slow ($\alpha \gg 1$) limits of the Kubo-Anderson model from \autoref{eqA:phi} result in 
\begin{equation}\label{eqA:philimits}
\begin{split}
    \phi(t) = 
\begin{cases}
    e^{-\Gamma t}, \quad\quad \text{fast limit}, \Gamma=\tau_c \Omega^2 \\
    e^{-\frac{1}{2}\Omega^2t^2}, \quad \text{slow limit}
\end{cases}
\end{split}
\end{equation}
The spectrum may now be calculated analytically with the last two terms in \autoref{eqA:dEdt} together with \autoref{eqA:barafinal}.

In the fast limit, the spectrum is given by
\begin{equation}\label{eqA:Iw-nc-fast}
\begin{split}
    I(\omega) 
    &= |F|^2 \Bigg(
    \int_0^\infty dt \, 
    e^{ i\bar{\omega}_0 t- \gamma t} e^{-\Gamma t} \\
    &+ \int_0^\infty dt \, 
    e^{ -i\bar{\omega}_0 t- \gamma t} e^{-\Gamma t}
    \Bigg) \\
    &= |F|^2 \frac{2(\gamma + \Gamma)}{\bar{\omega}_0^2 + (\gamma + \Gamma)^2}
\end{split}
\end{equation}
which reduces to Kubo-Anderson's result (when $\gamma=0$): a Lorentzian lineshape with width $\Gamma$ representing the case of homogeneous broadening.
In the slow limit the spectrum is given by
\begin{equation}
\begin{split}
    I(\omega) 
    &= |F|^2 \Bigg(
    \int_0^\infty dt \, 
    e^{ i\bar{\omega}_0 t- \gamma t} e^{-\frac{1}{2}\Omega^2 t^2}  \\
    &+
    \int_0^\infty dt \, 
    e^{ -i\bar{\omega}_0 t- \gamma t} e^{-\frac{1}{2}\Omega^2 t^2}
    \Bigg) \\
    &= \frac{1}{2}|F|^2\sqrt{\frac{2\pi}{\Omega^2}} \Bigg(
    \text{Erfc}\Big[\frac{(-i\bar{\omega}_0 + \gamma)}{\sqrt{2\Omega^2}}\Big]
    e^{\frac{1}{2\Omega^2}(-i\bar{\omega}_0 + \gamma)^2} \\
    &+ \text{Erfc}\Big[\frac{(i\bar{\omega}_0 + \gamma)}{\sqrt{2\Omega^2}}\Big]
    e^{\frac{1}{2\Omega^2}(i\bar{\omega}_0 + \gamma)^2}
    \Bigg)
\end{split}
\end{equation}
where $\text{Erfc}$ is the complementary error function.
By setting $\gamma=0$ we recover Kubo-Anderson's result of a Gaussian lineshape representing the case of inhomogeneous broadening.
\begin{equation}\label{eqA:Iw-nc-slow}
    I(\omega) =|F|^2\sqrt{\frac{2\pi}{\Omega^2}} 
    e^{-\frac{\bar{\omega}_0^2}{2\Omega^2}}
\end{equation}

%===================================================================================
\section{Analytical solution for the slow limit}
\begin{figure}
\includegraphics[width=70mm]{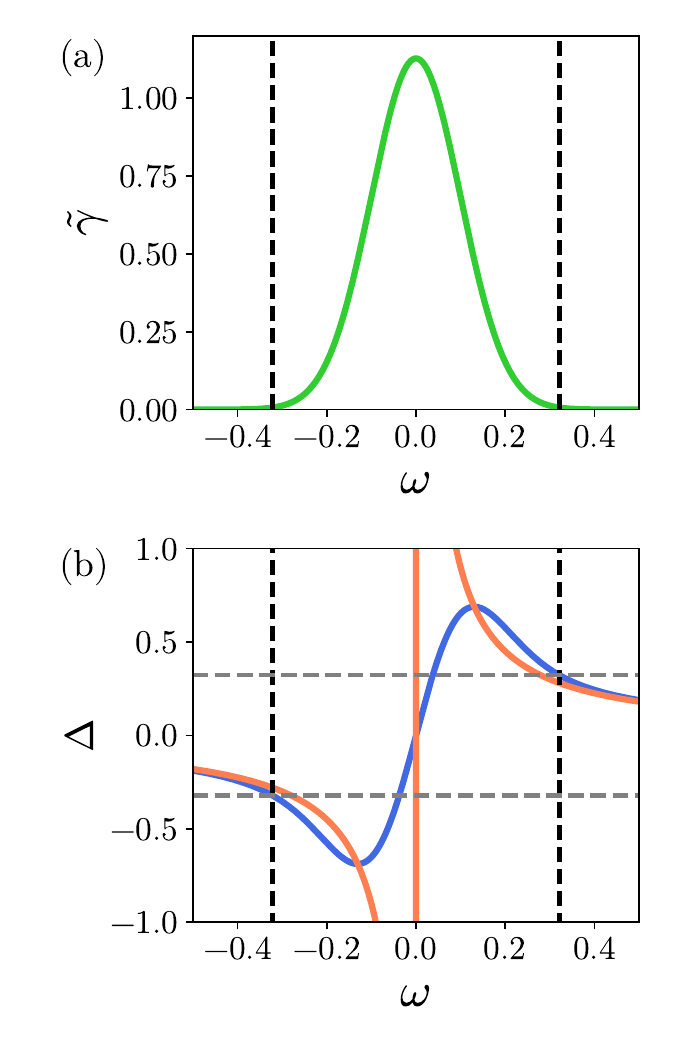}
\caption{(a) $\tilde{\gamma}$ and (b) $\Delta$ parameters as a function of the driving frequency. The blue line in (b) corresponds to the exact value while the red line corresponds to the approximate one given in \autoref{eqA:delta-approx}. The dashed (black and gray) lines indicate the polariton frequencies.
Parameters: $\omega_0=0$, $\kappa=0.02$, $\gamma=0$, $\Omega=0.1$, $\alpha=50$, and $\Omega_R=0.6$.}\label{fig-slow-functions}
\end{figure}

Using \autoref{eqA:philimits} in \autoref{eq:ac} leads to 
\begin{equation}\label{eqA:acslowsubs}
\begin{split}
    \langle \bar{a}_c \rangle  
    &= \frac{iF}{i\bar{\omega}_0 + \kappa + Nu^2\sqrt{\frac{\pi}{2\Omega^2}}
    \Big(1-i\text{Erfi}\big[\frac{\bar{\omega}_0 -i\gamma}{\sqrt{{2\Omega^2}}}\big] \Big) 
    e^{-\frac{(\bar{\omega}_0-i\gamma)^2}{2\Omega^2}}}
\end{split}
\end{equation}
where $\text{Erfi}$ denotes the imaginary error function. 
For simplicity we focus on the case where the molecules don't have any homogeneous broadening $(\gamma=0)$.
By defining 
$ \tilde{\gamma}(\omega) \equiv Nu^2\sqrt{\frac{\pi}{2\Omega^2}} e^{-\frac{\bar{\omega}_0^2}{2\Omega^2}} $ 
and 
$ \Delta(\omega) \equiv - \tilde{\gamma} \text{Erfi}\big[ \frac{\bar{\omega}_0}{\sqrt{2\Omega^2}} \big] $, 
the spectrum in the slow limit can be expressed as a Lorentzian profile
\begin{equation}\label{eqA:Iwslow}
    I(\omega) = |F|^2 \frac{2(\kappa + \tilde{\gamma})}
    {(\bar{\omega}_0 + \Delta)^2 +
    (\kappa + \tilde{\gamma})^2}
\end{equation}
In \autoref{fig-slow-functions} we plot $\tilde{\gamma}$ and $\Delta$ as a function of the driving frequency $\omega$ to gain some intuition on the meaning of these parameters, with the dashed lines indicating the polariton frequencies. We do this for $\omega_0=0$, $\kappa=0.02$, $\alpha=50$, $\Omega=0.1$ and $\Omega_R=0.6$. We see that $\tilde{\gamma}$ is a Gaussian function centered at $\omega_0$ with a variance of $\Omega$, which is the amplitude of the frequency modulations. When the Rabi splitting is much larger than $\Omega$, $\tilde{\gamma}$ does not contribute at the polariton frequencies, as easily seen in \autoref{fig-slow-functions}(a), and therefore, the width of the Lorentzian in \autoref{eqA:Iwslow} will be solely given by the cavity resonance broadening $\kappa$. Regarding the $\Delta$ parameter which determines the frequency of the poles of \autoref{eqA:Iwslow}, as we can see in \autoref{fig-slow-functions}(b) (blue line), when the driving frequency coincides with the polariton frequencies (dashed vertical black lines), $\Delta(\omega)$ exactly corresponds to this value.

We now aim to find an expression for the spectrum in \autoref{eqA:Iwslow} in the limit where the Rabi splitting is much larger than the amplitude of the frequency modulations. 
Recall that $\bar{\omega}_0 = \omega_0 - \omega$ and that we are in the large $N$ limit. At the polariton frequencies, $\bar{\omega}_0 \sim \pm \sqrt{N}u$, so by expanding $\Delta$ and $\tilde{\gamma}$ for 
$\sqrt{N}u / \sqrt{2}\Omega \gg 1$ we find that 
\begin{equation}\label{eqA:delta-approx}
    \Delta (\omega) = 
    -\frac{Nu^2}{\bar{\omega}_0} + O\Big(\frac{1}{\sqrt{N}}\Big) + i\tilde{\gamma}
\end{equation}
and $\tilde{\gamma} \rightarrow 0$. 
Note that the $1/\sqrt{N}$ scaling is in agreement with the result we obtained in \autoref{eq:scaling}. 

In \autoref{fig-slow-functions}(b) we plot this approximate $\Delta$ (red line) and verify that at the polariton frequencies, its value is very close to the exact one (blue line). Within this limit, the spectrum is then given by the following Lorentzian (which is consistent with \autoref{eq:Iwslowapprox})
\begin{equation}\label{eqA:Iwslowapprox}
    I(\omega) = |F|^2 \frac{2\kappa}{ 
    \Big(\bar{\omega}_0 - \dfrac{N u^2}{\bar{\omega}_0} \Big)^2
    + \kappa^2}
\end{equation}
In the strong coupling regime ($\sqrt{N}u \gg \kappa/2$), the poles of this expression are located at $\omega = \omega_0 \pm \sqrt{N}u -i\kappa/2$ and correspond to the two polariton peaks.

\begin{figure}
\includegraphics[width=\linewidth]{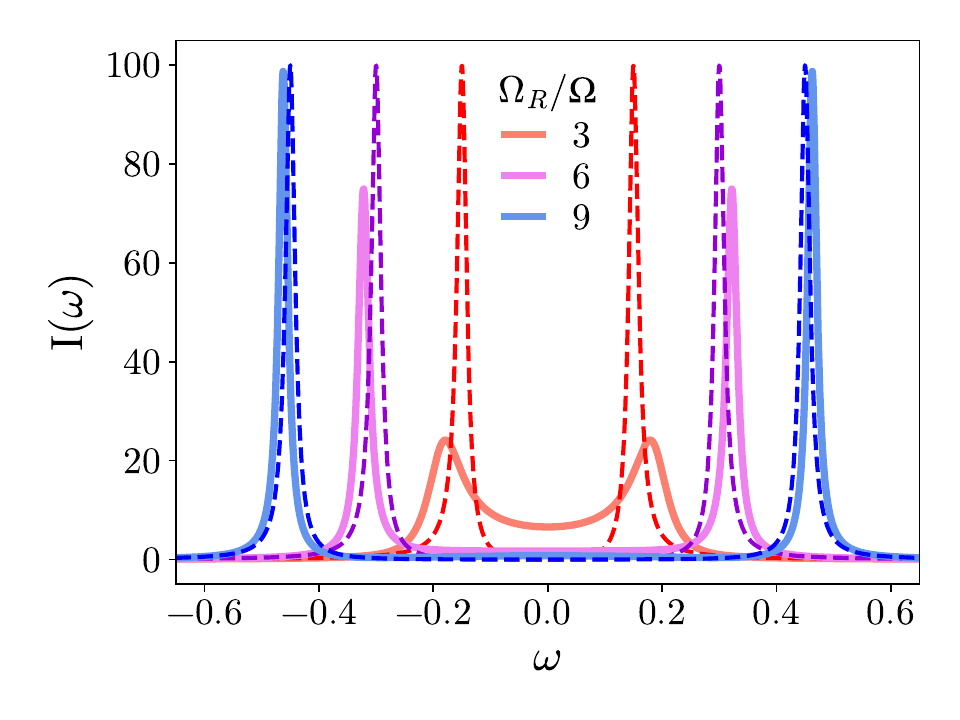}
\caption{Analytical spectrum for the slow modulation limit calculated with \autoref{eqA:Iwslow} (full lines) and \autoref{eqA:Iwslowapprox} (dashed lines). Note that as the Rabi splitting increases relative to the inhomogeneous broadening, the approximate spectra converge to the exact one.
Parameters: $\omega_0=0$, $\kappa=0.02$, $\Omega=0.1$, $\alpha=50$.}\label{fig-slow-approx}
\end{figure}

In \autoref{fig-slow-approx} we plot the spectrum in \autoref{eqA:Iwslow} (full lines) and compare it to the approximate one in \autoref{eqA:Iwslowapprox} (dashed lines) for increasing ratio of the Rabi splitting relative to the amplitude of the random frequency modulations. We observe that as $\Omega_R/\Omega$ increases, both spectra converge. 
Moreover, our results indicate that the Kubo-Anderson model is able to capture analytically the experimental observation that polaritons do not acquire the original inhomogeneous broadening of the molecular transitions. We note that this observation has previously been investigated numerically by several authors and analytically via the Green's function \cite{1996-houdre,1996-yamamoto, tgera1}. 

If we consider the general case where apart from static disorder, the molecular transitions also possess some homogeneous broadening $\gamma$, the spectral lineshape can readily be obtained from \autoref{eq:dEdt} and \autoref{eqA:acslowsubs}. We note here that in the strong coupling regime, the poles are given by \autoref{eq:poles}, replacing $\gamma_m$ by $\gamma$. 

%===================================================================================

\end{document}